\documentclass[baaa]{baaa}

\usepackage[pdftex]{hyperref}
\usepackage{subfigure}
\usepackage{natbib}
\usepackage{helvet,soul}
\usepackage[font=small]{caption}

\begin{document}


\journalvol{61A}
\journalyear{2019}
\journaleditors{R. Gamen, N. Padilla, C. Parisi, F. Iglesias \& M. Sgr\'o}


\contriblanguage{0}


\contribtype{2}

\thematicarea{1}

\title{Galaxias de bajo brillo superficial: análogas a los satélites de Andrómeda en Pegasus I?}
%
%
%
\titlerunning{Galaxias de bajo brillo superficial en Pegasus\,I}
\author{N.\,Gonz\'alez\inst{1,2,3},
        A.\,Smith\,Castelli\inst{1,2,3},
        F.\,Faifer\inst{1,2,3},
	C.\,Escudero\inst{1,2,3},
        S.\,A.\,Cellone\inst{1,3,4},
}
\authorrunning{González et al.}
%
%
\contact{ngonzalez@fcaglp.unlp.edu.ar}
\institute{Facultad de Ciencias Astron\'omicas y Geof\'isicas,  
        UNLP, La Plata, Argentina.
        \and  
        Instituto de Astrof\'isica de La  Plata (CCT La Plata, CONICET
        - UNLP), La Plata, Argentina.
        \and 
        Consejo   Nacional   de    Investigaciones   Cient\'ificas   y
        T\'ecnicas, Argentina.
        \and
        CASLEO, San Juan, Argentina.
}
%
%
\resumen{En este trabajo presentamos los resultados preliminares de un
  estudio   donde  parece   existir  un   efecto  de   sesgo  en   las
  distribuciones de tamaño de las  galaxias de bajo brillo superficial
  (LSB) detectadas  en diferentes entornos,  en el sentido de  que los
  grupos/cúmulos más  distantes carecen  de objetos de  radio efectivo
  pequeños, mientras que  los sistemas grandes no se  encuentran en el
  Grupo  Local y  los  entornos  cercanos.  Si  bien  puede haber  una
  escasez real  de galaxias LSB  grandes en entornos de  baja densidad
  como  el  Grupo Local,  la  no  detección  de sistemas  pequeños  (y
  débiles) a grandes distancias es  claramente un efecto de selección.
  Como ejemplo,  las galaxias LSB con  tamaños similares a las  de los
  satélites de Andrómeda en el Grupo Local, ciertamente se perderán en
  una identificación visual a la distancia de Pegasus\,I.}
\abstract{Here we show the preliminary  results of a study where there
  seems to be a bias effect  in the size distributions of the detected
  low-surface brightness (LSB) galaxies in different environments.  In
  this sense, more distant  groups/clusters would lack small effective
  radius objects,  while large  systems would not  found in  the Local
  Group  and  nearby  environments.   While there  may  be  an  actual
  shortage of large LSB galaxies  in low-density environments like the
  Local Group, the non-detection of small (and faint) systems at large
  distances is clearly a selection effect. As an example, LSB galaxies
  with similar  sizes to those of  the satellites of Andromeda  in the
  Local Group, will be certainly  missed in a visual identification at
  the distance of Pegasus\,I.}
%
%
\keywords{Galaxies: dwarf ---  galaxies:  photometry  ---  galaxies:
  groups: (Pegasus\,I)}
\maketitle
\section{Introducción}
\label{S_intro}
En el  trabajo de  \citet{2018arXiv181000710G} se plantea  una posible
correlación entre  la distancia de los  cúmulos/grupos que albergarían
galaxias de bajo brillo superficial (LSB) y sus radios efectivos. Esta
correlación parece indicar que las  galaxias LSB poseen un tamaño real
más grande  a mayores  distancias.  En  cambio, sus  tamaños aparentes
resultan más pequeños a distancia  lejanas. Este efecto podría deberse
solamente a las  limitaciones que resultan de observar  estos tipos de
objetos  de  bajo  brillo  superficial. Esto  es,  las  LSB  pequeñas
observadas  a grandes  distancias podrían  confundirse con  objetos de
fondo,  mientras  que  aquellas   galaxias  LSB  cercanas  podrían  no
detectarse debido a sus  tamaños angulares extremadamente grandes. Sin
embargo,  la probabilidad  de  encontrar galaxias  LSB  grandes en  un
volumen    pequeño    alrededor    del    Grupo    Local    resultaría
baja. Recientemente, \citet{Muller2018}  encontraron posibles galaxias
ultra  difusas (UDGs)  extremadamente  grandes en  el  grupo de  Leo-I
(D~$\sim$~10.7~Mpc). De confirmarse su pertenencia en este grupo, estos
objetos pasarán a ser las UDGs más cercanas. 
%
\begin{figure}[!t]
  \centering
  \includegraphics[width=0.41\textwidth]{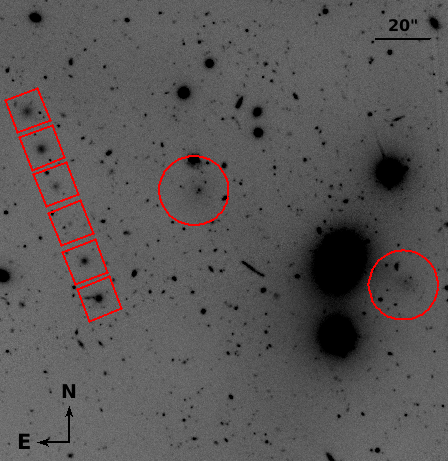}
  \caption{Mosaico de  2.85~$\times$~2.85~arcmin en el  filtro $g'$ en
    el  cual  se  muestra  dos  de  las  galaxias  LSB  detectadas  en
    Pegasus\,I  (círculos) reportadas  en \citet{2018arXiv181000710G}.
    Superpuesto en este  mosaico, a modo de comparación  se muestra la
    apariencia  que tendrían  las dSphs  de Andrómeda  llevadas a  la
    distancia de Pegasus\,I (cuadrados).}
  \label{Figura1}
\end{figure}
\begin{figure*}[!t]
  \centering
  \includegraphics[width=0.87\textwidth]{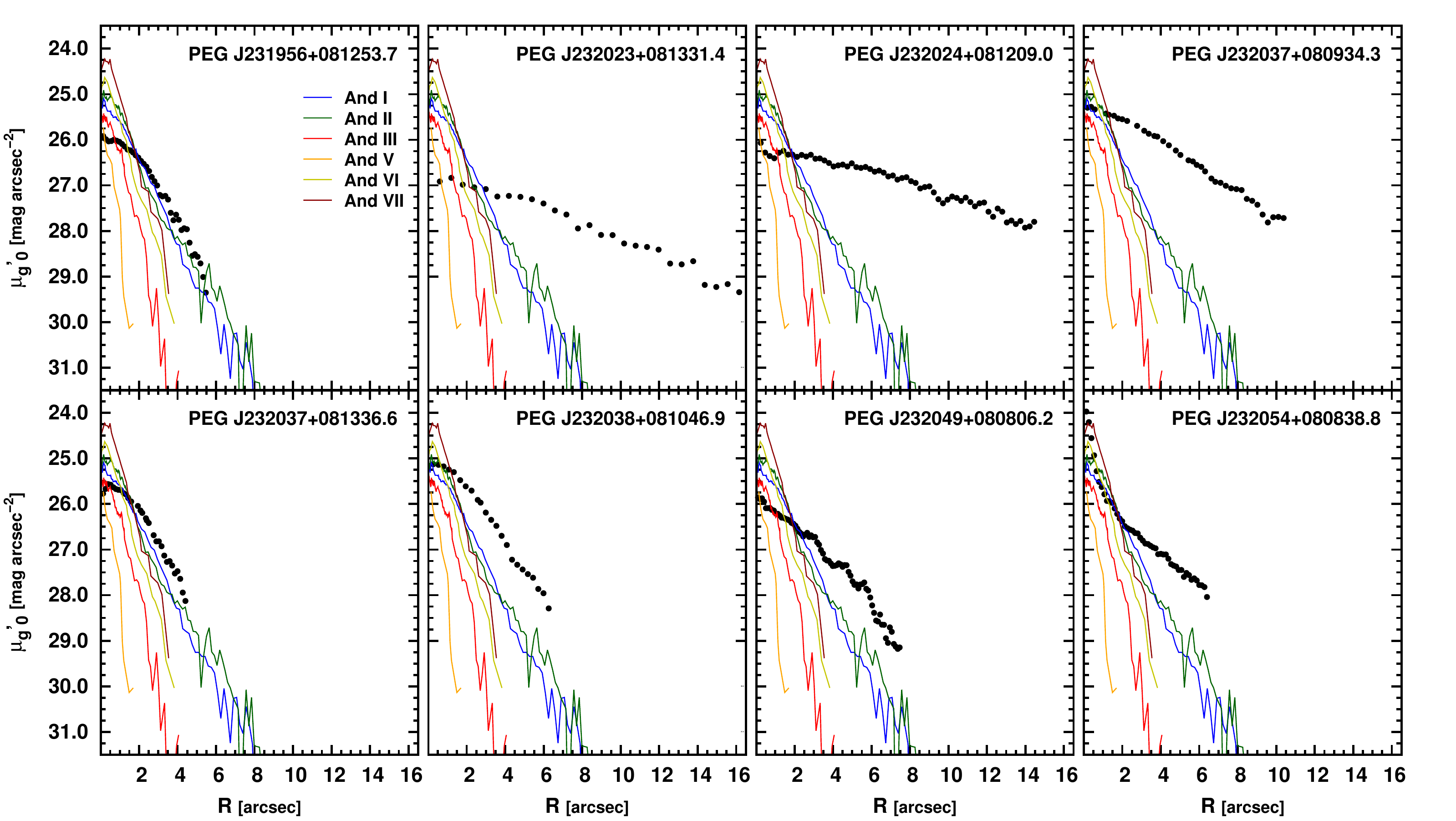}
  \caption{Comparación  entre  los  perfiles  de brillo  $g'$  de  las
    galaxias LSB detectadas en  Pegasus I (círculos negros) reportadas
    en \citet{2018arXiv181000710G} y las  seis galaxias dSph satélites
    de Andrómeda \citep{1992AJ....103..840C,1999AJ....118.1230C}.  Los
    perfiles de  brillo en  la banda  V de  las galaxias  satélites de
    Andrómeda fueron  transformados a  la banda $g'$  a través  de las
    relaciones  presentadas  por   \citet{1995PASP..107..945F},  y  se
    escalearon a la distancia de Pegasus\,I.}
  \label{Figura2}
\end{figure*}
\begin{figure}[!t]
  \centering
  \includegraphics[width=0.405\textwidth]{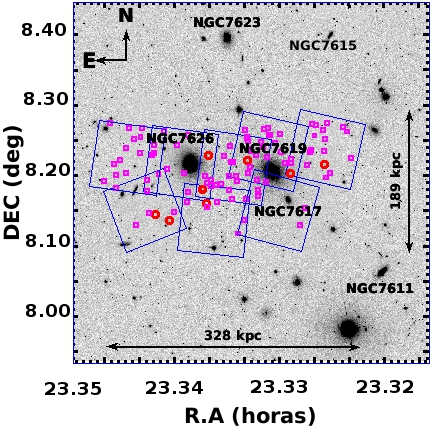}
  \caption{Mosaico  de 30.5$\times$30.5~arcmin  en el  filtro $r'$  de
    SDSS DR12 mostrando la región central del grupo de Pegasus\,I. Los
    marcos azules corresponden a los campos de GEMINI- GMOS utilizados
    en este  trabajo ($\sim$~5.5~arcmin de lado).   Los círculos rojos
    indican     la     ubicación     de     las     candidatas     LSB
    \citep{2018arXiv181000710G}.   Los cuadrados  violetas indican  la
    ubicación   de  galaxias   candidatas  cuya   morfología  resultan
    similares a las galaxias satélites observadas en Andrómeda.}
  \label{Figura3}
\end{figure}
\begin{figure}[!t]
  \centering
  \includegraphics[width=0.24\textwidth]{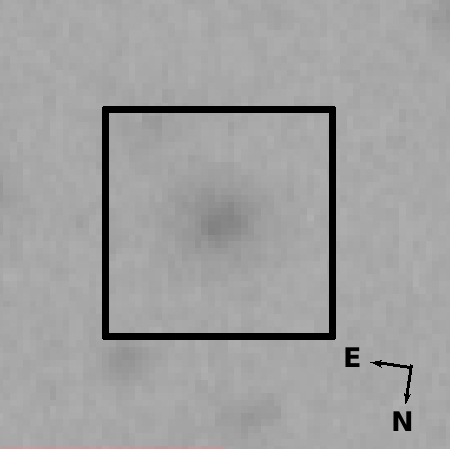}
  \includegraphics[width=0.24\textwidth]{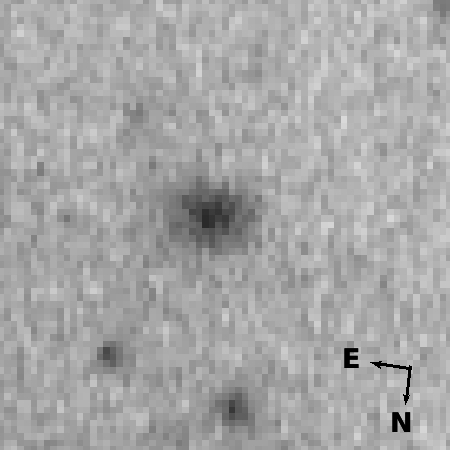}\\
  \vspace{0.1cm}
  
  \includegraphics[width=0.24\textwidth]{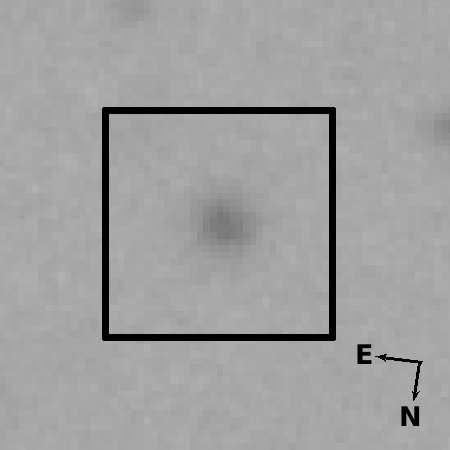}
  \includegraphics[width=0.24\textwidth]{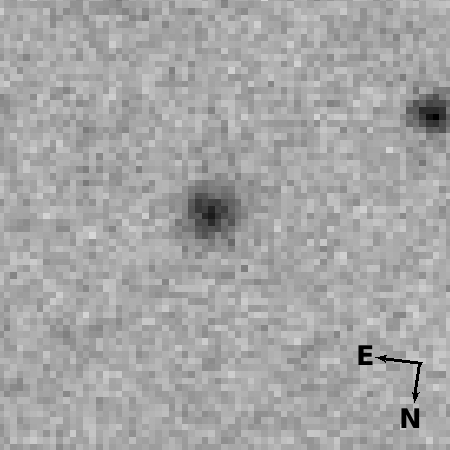}\\
  \vspace{0.09cm}
  
  \includegraphics[width=0.24\textwidth]{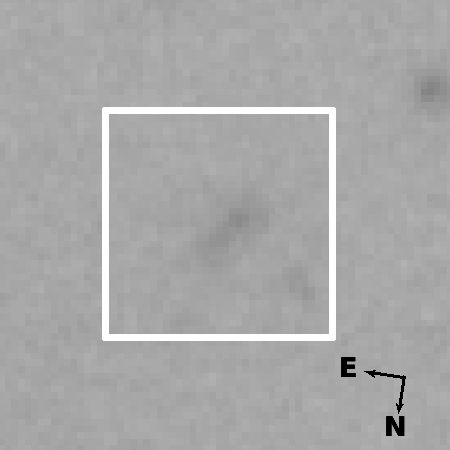}
  \includegraphics[width=0.24\textwidth]{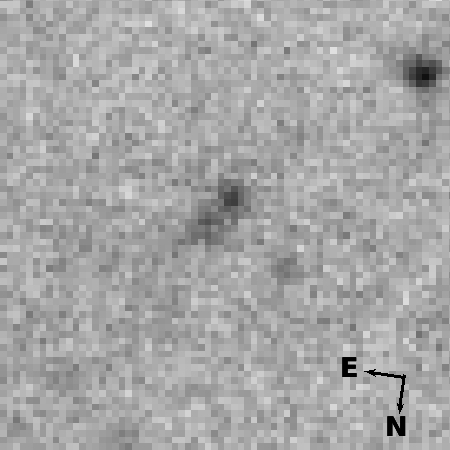}\\
  \vspace{0.09cm}
  
  \includegraphics[width=0.24\textwidth]{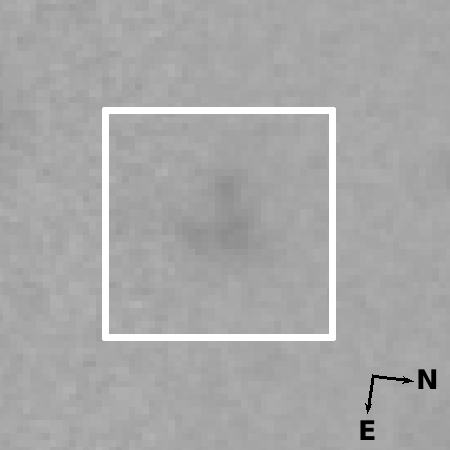}
  \includegraphics[width=0.24\textwidth]{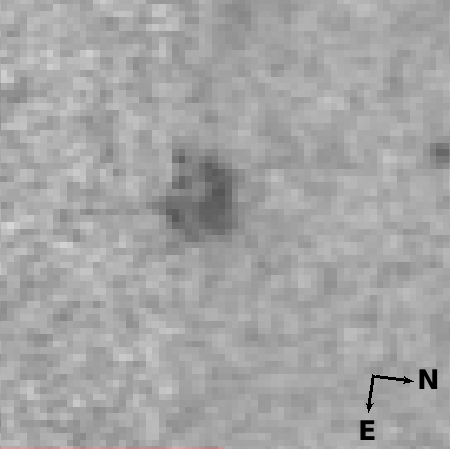}\\
  \caption{En  los  paneles izquierdos  se  muestran  las imágenes  de
    10$\times$10~arcsec en  el filtro  $g'$ de  cuatro de  los objetos
    encontrados en la primera búsqueda con características similares a
    los satélites  de Andrómeda a  la distacia de Pegasus\,I.   En los
    paneles derechos se muestran sus respectivas imágenes en el filtro
    $i'$.   En  ambos  filtros  se considero  los  mismos  niveles  de
    despliegue.  Los cuadros negros señalan los objetos petencialmente
    canditados, los  cuadros blancos  indican los objetos  que podrían
    ser descartados.}
    \label{Figura4}
\end{figure}
\section{Selecci\'on de las candidatas a LSB}
\label{S_seleccion}
Con respecto  a la posible  presencia de un  sesgo en la  detección de
galaxias  LSB,  nos  preguntamos   cuántos  objetos  similares  a  los
satélites  dSph de  Andrómeda ($D~\sim~0.784$~Mpc)  aparecerían a  la
distancia de  Pegasus\,I ($D~\sim~50$~Mpc).  En  la Fig.~\ref{Figura1}
mostramos  los modelos  de seis  de  estás galaxias  obtenidas de  los
perfiles  de   brillo  reportados  por   \cite{1992AJ....103..840C}  y
\cite{1999AJ....118.1230C},   sobrepuesto  en   uno   de  los   campos
GEMINI-GMOS en Pegasus\,I.

Adicionalmente, comparamos  los perfiles de brillo  superficial de las
galaxias LSB  con los de  los satélites  de Andrómeda escaleados  a la
distancia de  Pegasus\,I (ver  Fig.~\ref{Figura2}).  Se puede  ver que
las  dSphs de  Andrómeda  muestran  tamaños entre  $2~\textless~r_{\rm
  tot}~\textless~8$~arcsec  y  brillos superficiales  centrales  entre
$24.5~\textless~\mu_{0,g'}~\textless~26$    mag~arcsec${^{-2}}$.    Su
brillo superficial  central es bastante  bajo, pero aún  más brillante
que los de las galaxias LSB  detectadas en Pegasus\,I. Sin embargo, en
comparación, sus tamaños aparentes resultan más pequeños.

En  este contexto,  la pregunta  que surge  es, ¿Cuántas  LSB de  tipo
temprano  se podrían  perder en  la inspección  visual porque  son muy
pequeñas y fáciles de confundir con los objetos de fondo?  Para buscar
estos    objetos   decidimos    utilizar   el    software   SExtractor
\citep{1996A&AS..117..393B} adoptando los siguientes criterios:
\begin{align*} 
  &{\rm CLASS\_STAR} \leqslant0.2, \\
  &{\rm FLAGS} \leqslant 2, \\
  &{\rm Elipticidad} \leqslant 0.2,\\
  &2~\textless~r_{\rm tot}~\textless~8~{\rm arcsec},\\
  &24.5 \lesssim \mu_{0_{(g',r',i')}} \lesssim~26~{\rm mag~arcsec}^{-2},\\
  &0~\lesssim~g'-i'~\lesssim~1.3,\\
  &{\rm y}~0~\lesssim~g'-r'~\lesssim~1.2\\
\end{align*}
Se  adoptaron  estos  rangos   de  colores  considerando  los  valores
mostrados para las galaxias LSB de  tipo temprano en la sección 5.1 de
\citet{2018arXiv181000710G}.
%
%
\section{Datos fotométricos}
Este  trabajo está  basado en  ocho  campos profundos  tomados en  los
filtros $g'$,  $r'$ e  $i'$ \citep{1996AJ....111.1748F},  empleando la
cámara GMOS  de Gemini Norte  (Programa GN-2008B-Q-14, PI:  F. Faifer;
Programa   GN-2012A-Q-55,    PI:   A.    Smith    Castelli;   Programa
GN-2012B-Q-69,  PI:  F.   Faifer  ;  Programa  GN-2014A-Q-70,  PI:  F.
Faifer;   Programa   GN-2014B-Q-17,   PI:   N.    González;   Programa
GN-2015B-Q-13, PI:  N.  González).   Estas imágenes cubren  el entorno
cercano  a las  dos  galaxias dominantes  de  Pegasus\,I: NGC\,7626  y
NCG\,7619, y se utilizaron para  obtener la selección preliminar de de
noventa galaxias  candidatas cuya morfología resultan  similares a las
galaxias  satélites de  Andrómeda.  La  Fig.~\ref{Figura3} muestra  la
orientación  de los  diferentes campos  analizados y  la ubicación  de
estos objetos. Por su parte, en la Fig.~\ref{Figura4}, se muestran las
imágenes de cuatro de estás candidatas en los filtros $g'$ e $i'$.

Dado que las galaxias LSB con tamaños similares a las de los satélites
de Andrómeda serán difíciles de detectar en una inspección visual a la
distancia de  Pegasus\,I, la identificación  de las mismas  se realizó
utilizando  SExtractor  en  las   imágenes  GEMINI-GMOS.   Para  ello,
utilizando  las  tareas {\it  ellipse}  y  {\it  bmodel} de  IRAF,  se
procedió a  modelar las  distribuciones de  brillo superficial  de las
galaxias  elípticas NGC\,7626  y  NGC\,7619 y  sus respectivos  halos,
incluyendo  varios  objetos  extendidos.  Luego,  se  restaron  dichos
modelos  con el  fin de  poder identificar  y medir  las candidatas  a
galaxias LSB (para  más detalles, remitimos al lector  a la sección\,3
de   \citealt{2018arXiv181000710G}).   Posteriormente   realizamos  la
fotometría  con   SExtractor  utilizando  el  criterio   de  selección
mencionado en la Sec.~\ref{S_seleccion},  el cual permitió identificar
noventa  objetos  con características  similares  a  los satélites  de
Andrómeda a la distancia de Pegasus\,I.
\section{Resultados preliminares y trabajo a futuro}
En este  trabajos se realizó la  detección de objetos similares  a las
galaxias satélites  dSph de  Andrómeda en  el grupo  Pegasus\,I.  Como
resultado,  se  detectaron  en  el grupo  noventa  objetos  con  estas
características. El primer paso,  será realizar una revisión detallada
de estos  objetos, y analizar  si los  mismos presentan algún  tipo de
subestructura y/o  posible formación estelar,  ya que no  es esperable
que  estas  características  esten   presentes  en  galaxias  de  tipo
temprano. Para ello, se comenzará  inspeccionando las imágenes de cada
objeto   en  los   filtros  $g'$   e  $i'$.    Como  ejemplo,   en  la
Fig.~\ref{Figura4}, se muestran en  los paneles superiores dos objetos
que  son potencialmente  candidatos  a ser  galaxias  similares a  las
satélites de Andrómeda.  En contraparte,  en los paneles inferiores se
muetran  dos  objetos  que  podrían  ser  descartados  de  la  primera
selección por mostrar un cierta subestructura.
\bibliographystyle{baaa}
\small
\bibliography{biblio}
\end{document}